# PERFORMANCE COMPARISON OF TWO CLIPPING BASED FILTERING METHODS FOR PAPR REDUCTION IN OFDM SIGNAL

Md. Munjure Mowla[1], Md. Yeakub Ali[2] and Rifat Ahmmed Aoni[3]

[1,2]Department of Electronics & Telecommunication Engineering,
Rajshahi University of Engineering & Technology, Rajshahi - 6204, Bangladesh
[3]Department of Electrical Engineering,
University of Malaya, Kuala Lumpur -50603, Malaysia

## ABSTRACT

*The growth of wireless communication technologies has been producing the intense demand for high-speed, efficient, reliable voice & data communication. As a result, third generation partnership project (3GPP) has implemented next generation wireless communication technology long term evolution (LTE) which is designed to increase the capacity and speed of existing mobile telephone & data networks. LTE has adopted a multicarrier transmission technique known as orthogonal frequency division multiplexing (OFDM). OFDM meets the LTE requirement for spectrum flexibility and enables cost-efficient solutions for very wide carriers. One major generic problem of OFDM technique is high peak to average power ratio (PAPR) which is defined as the ratio of the peak power to the average power of the OFDM signal. A trade-off is necessary for reducing PAPR with increasing bit error rate (BER), computational complexity or data rate loss etc. In this paper, two clipping based filtering methods have been implemented & also analyzed their modulation effects on reducing PAPR.*

## KEYWORDS

*Bit Error rate (BER), Complementary Cumulative Distribution Function (CCDF), Long Term Evolution (LTE), Orthogonal Frequency Division Multiplexing (OFDM) and Peak to Average Power Ratio (PAPR).*

## 1. INTRODUCTION

Orthogonal frequency division multiplexing (OFDM) is a multicarrier modulation (MCM) technique which seems to be an attractive candidate for fourth generation (4G) wireless communication systems. The additional increasing demand on high data rates in wireless communications systems has arisen in order to carry broadband services. OFDM offers high spectral efficiency, immune to the multipath fading, low inter-symbol interference (ISI), immunity to frequency selective fading and high power efficiency. OFDM has been adopted by Digital Audio Broadcasting (DAB), Digital Video Broadcasting-Terrestrial (DVB-T) and Wireless LAN (WLAN) systems. Additionally, OFDM has been used in the mobility mode of IEEE802.16 WiMAX. Furthermore, it is currently a working specifications in 3GPP Long Term Evolution (LTE) downlink, and is the candidate access method for the IEEE 802.22 Wireless Regional Area Networks (WRAN) [1].

One of the major problems of OFDM is high peak to average power ratio (PAPR) of the transmit signal. If the peak transmit power is limited by either regulatory or application constraints, the effect is to reduce the average power allowed under multicarrier transmission relative to that under constant power modulation techniques. This lessens the range of multicarrier transmission. Furthermore, the transmit power amplifier must be operated in its linear region (i.e., with a large input back-off), where the power conversion is inefficient to avoid spectral growth of the multicarrier signal in the form of intermodulation among subcarriers and out-of-band radiation. This may have a deleterious effect on battery lifetime in mobile applications. As handy devices have a finite battery life it is significant to find ways of reducing the PAPR allowing for a smaller more efficient high power amplifier (HPA), which in turn will mean a longer lasting battery life. In many low-cost applications, the problem of high PAPR may outweigh all the potential benefits of multicarrier transmission systems [2].

A number of promising approaches or processes have been proposed & implemented to reduce PAPR with the expense of increase transmit signal power, bit error rate (BER) & computational complexity and loss of data rate, etc.  So, a system trade-off is required. These techniques include Amplitude Clipping and Filtering, Peak Windowing, Peak Cancellation, Peak Reduction Carrier, Envelope Scaling, Decision-Aided Reconstruction (DAR), Coding, Partial Transmit Sequence (PTS), Selective Mapping (SLM), Interleaving, Tone Reservation (TR), Tone Injection (TI), Active Constellation Extension (ACE), Clustered OFDM, Pilot Symbol Assisted Modulation, Nonlinear Companding Transforms etc [3].

## 2. THEORETICAL MODEL OF OFDM SYSTEM

OFDM is a special form of multicarrier modulation (MCM) with densely spaced subcarriers with overlapping spectra, thus allowing multiple-access [4]. MCM works on the principle of transmitting data by dividing the stream into several bit streams, each of which has a much lower bit rate and by using these sub-streams to modulate several carriers.

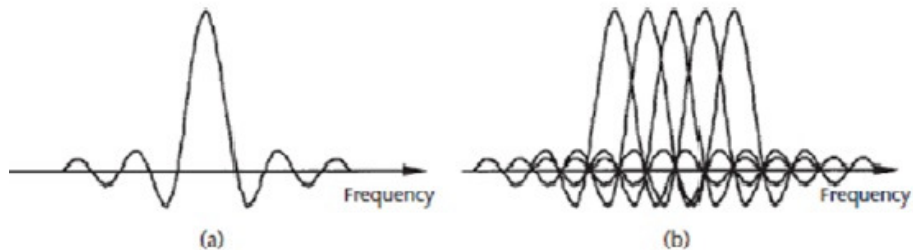

Figure 1.   Spectra of (a) An OFDM Sub-channel and (b) An OFDM Signal [1]

In multicarrier transmission, bandwidth divided in many non-overlapping subcarriers but not necessary that all subcarriers are orthogonal to each other as shown in figure 1 (a) [4]. In OFDM the sub-channels overlap each other to a certain extent as can be seen in figure 1 (b), which leads to an efficient use of the total bandwidth. The information sequence is mapped into symbols, which are distributed and sent over the N sub-channels, one symbol per channel. To permit dense packing and still ensure that a minimum of interference between the sub-channels is encountered, the carrier frequencies must be chosen carefully.  By using orthogonal carriers, frequency domain can be viewed so as the frequency space between two sub-carriers is given by the distance to the first spectral null [1].

### 2.1. Mathematical Explanation of OFDM Signals

In OFDM systems, a defined number of successive input data samples are modulated first (e.g, PSK or QAM), and then jointly correlated together using inverse fast Fourier transform (IFFT) at the transmitter side. IFFT is used to produce orthogonal data subcarriers. Let, data block of length $N$ is represented by a vector, $X=[X_0, X_1,...... X_{N-1}]^T$. Duration of any symbol $X_K$ in the set $X$ is $T$ and represents one of the sub-carriers set. As the N sub-carriers chosen to transmit the signal are orthogonal, so we can have, $f_n = n\Delta f$, where $n\Delta f = 1/NT$ and $NT$ is the duration of the OFDM data block $X$. The complex data block for the OFDM signal to be transmitted is given by [2],

$$x(t) = \frac{1}{\sqrt{N}} \sum_{n=0}^{N-1} X_n e^{j2\pi n\Delta ft} \qquad 0 \leq t \leq NT \qquad (1)$$

Where,

$j = \sqrt{-1}$ , $\Delta f$ is the subcarrier spacing and $NT$ denotes the useful data block period. In OFDM the subcarriers are chosen to be orthogonal (i.e., $\Delta f = 1/NT$). However, OFDM output symbols typically have large dynamic envelope range due to the superposition process performed at the IFFT stage in the transmitter.

## 3. OVERVIEW OF PAPR

Presence of large number of independently modulated sub-carriers in an OFDM system the peak value of the system can be very high as compared to the average of the whole system. Coherent addition of N signals of same phase produces a peak which is N times the average signal [2]. PAPR is widely used to evaluate this variation of the output envelope. PAPR is an important factor in the design of both high power amplifier (PA) and digital-to-analog (D/A) converter, for generating error-free (minimum errors) transmitted OFDM symbols. So, the ratio of peak power to average power is known as PAPR.

$$PAPR = \frac{Peak\_Power}{Average\_Power}$$

The PAPR of the transmitted signal is defined as [5],

$$PAPR[x(t)] = \frac{\max_{0 \leq t \leq NT} |x(t)|^2}{P_{av}} = \frac{\max_{0 \leq t \leq NT} |x(t)|^2}{\frac{1}{NT} \int_0^{NT} |x(t)|^2 \, dt} \qquad (2)$$

Where, $P_{av}$ is the average power of and it can be computed in the frequency domain because Inverse Fast Fourier Transform (IFFT) is a (scaled) unitary transformation.

To better estimated the PAPR of continuous time OFDM signals, the OFDM signals samples are obtained by $L$ times oversampling. $L$ times oversampled time domain samples are $LN$ point IFFT of the data block with $(L-1)N$ zero-padding. Therefore, the oversampled IFFT output can be expressed as [2],

$$x[n] = \frac{1}{\sqrt{N}} \sum_{k=0}^{N-1} X_k e^{j2\pi nk/LN} \qquad 0 \leq n \leq NL-1 \qquad (3)$$

It is known that the PAPR of the continuous-time signal cannot be obtained precisely by the use of Nyquist rate sampling, which corresponds to the case of L = 1. It is shown in that L = 4 can provide sufficiently accurate PAPR results.

The PAPR computed from the L-times oversampled time domain signal samples is given by[2],

$$PAPR\{x[n]\} = \frac{\max_{0 \leq t \leq NL-1} |x(n)|^2}{E[|x(n)|^2]} \qquad (4)$$

Where, E{.} is the Expectation Operator.

## 4. AMPLITUDE CLIPPING AND FILTERING

Amplitude Clipping and Filtering is one of the easiest techniques which may be under taken for PAPR reduction for an OFDM system. A threshold value of the amplitude is fixed in this case to limit the peak envelope of the input signal [5].

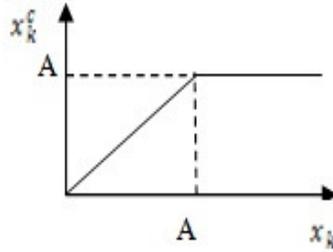

Figure 2. Clipping Function

The clipping ratio (CR) is defined as,

$$CR = \frac{A}{\sigma} \qquad (5)$$

Where, A is the amplitude and $\sigma$ is the root mean squared value of the unclipped OFDM signal. The clipping function is performed in digital time domain, before the D/A conversion and the process is described by the following expression,

$$x_k^c = \begin{cases} x_k & |x_k| \leq A \\ Ae^{j\phi(x_k)} & |x_k| > A \end{cases} \qquad 0 \leq k \leq N-1 \qquad (6)$$

Where, $x_k^c$ is the clipped signal, $x_k$ is the transmitted signal, A is the amplitude and $\phi(x_k)$ is the phase of the transmitted signal $x_k$.

### 4.1. Limitations of Clipping and Filtering

- ➢ Clipping causes in-band signal distortion, resulting in BER performance degradation.
- ➢ Clipping also causes out-of-band radiation, which imposes out-of-band interference signals to adjacent channels. Although the out-of-band signals caused by clipping can be reduced by filtering, it may affect high-frequency components of in-band signal (aliasing) when the clipping is performed with the Nyquist sampling rate in the discrete-time domain[6].
- ➢ However, if clipping is performed for the sufficiently-oversampled OFDM signals (e.g., L ≥4) in the discrete-time domain before a low-pass filter (LPF) and the signal passes through a band-pass filter (BPF), the BER performance will be less degraded [6].
- ➢ Filtering the clipped signal can reduce out-of-band radiation at the cost of peak regrowth. The signal after filtering operation may exceed the clipping level specified for the clipping operation [2].

## 5. PROPOSED CLIPPING AND FILTERING SCHEME

As the major spotlight of this research is to reduce PAPR, so, in this simulation, we have trade-off between PAPR reduction with BER increment. Very little amount of BER increment is desirable. Pointing out the third limitation in section 4.1, Our Previous Work [7] showed that if clipped signal passes through a Composed filter (FIR based HPF) before passing a LPF to reduce out-of-band radiation, then it causes less BER degradation with medium amount of PAPR reduction than an existing method [6]. Considering this concept, we have designed another scheme for clipping & filtering method where clipped signal passes through a Composed filter (IIR based BPF) before passing a LPF, then it causes a little bit more BER degradation but more amount of PAPR reduction than our previous work[7].

This proposed scheme is shown in the figure 4. It shows a block diagram of a PAPR reduction scheme using clipping and filtering, where $L$ is the oversampling factor and $N$ is the number of subcarriers. The input of the IFFT block is the interpolated signal introducing $N(L-1)$ zeros (also, known as zero padding) in the middle of the original signal is expressed as,

$$X'[k] = \begin{cases} X[k], & \text{for } 0 \leq k \leq \frac{N}{2} \text{ and } NL - \frac{N}{2} < k < NL \\ 0 & \text{elsewhere} \end{cases} \quad (7)$$

In this system, the L-times oversampled discrete-time signal is generated as,

$$x'[m] = \frac{1}{\sqrt{LN}} \sum_{k=0}^{L.N-1} X'[k] e^{\frac{j2\pi n \Delta f k}{L.N}}, \qquad m = 0,1,\ldots NL-1 \quad (8)$$

and is then modulated with carrier frequency $f_c$ to yield a passband signal $x^p[m]$.

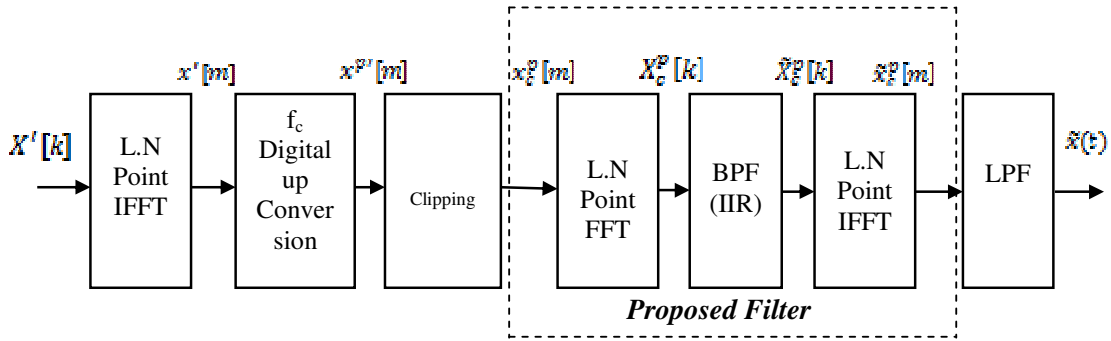

Figure 4. Block Diagram of Proposed Clipping & Filtering Scheme.

Let $x_c^p[m]$ denote the clipped version of $x^p[m]$, which is expressed as,

$$x_c^p[m] = \begin{cases} -A & x^p[m] \leq -A \\ x^p[m] & |x^p[m]| < A \\ A & x^p[m] \geq A \end{cases} \quad (9)$$

Or,

$$x_c^p[m] = \begin{cases} x^p[m] & if, \ |x^p[m]| < A \\ \frac{x^p[m]}{|x^p[m]|} \cdot A & otherwise \end{cases} \quad (10)$$

Where, *A* is the pre-specified clipping level. After clipping, the signals are passed through the Composed filter *(Proposed Filter).* This composed filter itself consists on a set of FFT-IFFT operations where filtering takes place in frequency domain after the FFT function. The FFT function transforms the clipped signal $x_c^p[m]$ to frequency domain yielding $X_c^p[k]$. The information components of $X_c^p[k]$ are passed to a band pass filter (BPF) producing $\tilde{X}_c^p[k]$. This filtered signal is passed to the unchanged condition of IFFT block and the out-of-band radiation that fell in the zeros is set back to zero. The IFFT block of the filter transforms the signal to time domain and thus obtain $\tilde{x}_c^p[m]$.

## 6. DESIGN AND SIMULATION PARAMETERS

In this simulation, an IIR digital filter (Chebyshev Type I) is used in the composed filtering. Chebyshev Type I filter is equiripple in the passband and monotonic in the stopband. Type I filter rolls off faster than type II filters. Chebyshev filter has the property that it minimizes the error between the idealized and the actual filter characteristic over the range of the filter. Because of the passband equiripple behaviour inherent in Chebyshev Type I filter, it has a smoother response.

Table 1 shows the values of parameters used in the QPSK & QAM system for analyzing the performance of clipping and filtering technique [6]. We have simulated the both methods (Previous and New) with the same parameters at first and compare each step.

Table 1. Parameters Used for Simulation of Clipping and Filtering.

| Parameters | Value |
|---|---|
| Bandwidth ( BW) | 1 MHz |
| Over sampling factor (L) | 8 |
| Sampling frequency, $f_s$ = BW*L | 8 MHz |
| Carrier frequency, $f_c$ | 2 MHz |
| FFT Size / No. of Subscribers (N) | 128 |
| CP / GI size | 32 |
| Modulation Format | QPSK and QAM |
| Clipping Ratio (CR) | 0.8, 1.0, 1.2, 1.4, 1.6 |

### 6.1. Simulation Results for PAPR Reduction

At first, we simulate the PAPR distribution for CR values =0.8, 1.0, 1.2, 1.4, 1.6 with QPSK modulation and N=128. Then, we simulate with QAM modulation and N=128 and compare different situations.

**6.1.1 Simulation Results: (QPSK and N=128)**

In the previous method, PAPR distribution for different CR value is shown in figure 5 (a). At CCDF =$10^{-1}$, the unclipped signal value is 13.51 dB and others values for different CR are tabulated in the table 2.

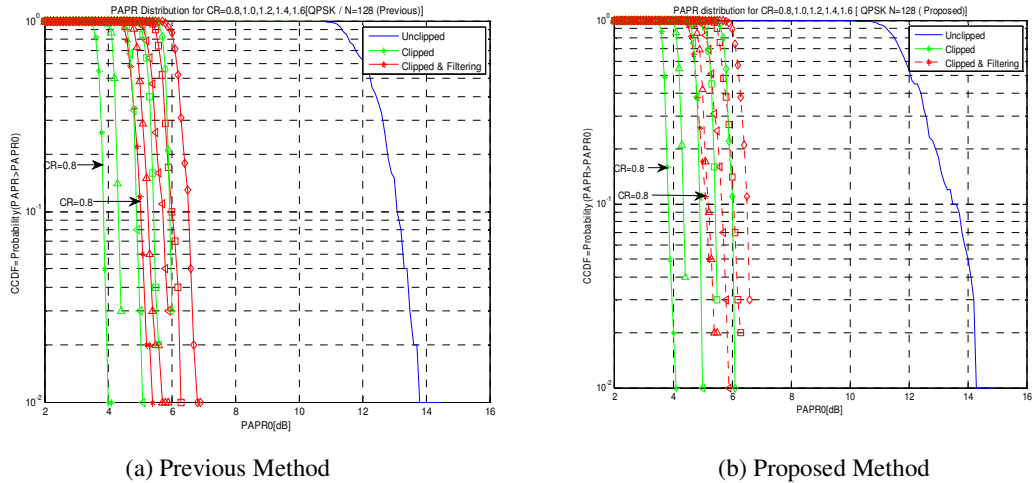

(a) Previous Method          (b) Proposed Method

Figure 5. PAPR Distribution for CR=0.8,1.0,1.2,1.4,1.6 [QPSK and N=128]

In the proposed method, simulation shows the reduction of PAPR is shown for different CR values in figure 5(b). At CCDF =$10^{-1}$, the unclipped signal value is 13.52 dB and others values for different CR are tabulated in the table 2. From table 2, it is clearly observed that the proposed method reduces PAPR with respect to previous work analysis [7].

Table 2. Comparison of Previous with Proposed Method for PAPR value [QPSK and N=128]

| CR value | PAPR value (dB) (Previous) | PAPR value (dB) (Proposed) | Improvement in PAPR value (dB) |
|---|---|---|---|
| 0.8 | 5.11 | 4.21 | 0.90 |
| 1.0 | 5.18 | 4.67 | 0.51 |
| 1.2 | 5.65 | 5.21 | 0.44 |
| 1.4 | 6.04 | 5.72 | 0.32 |
| 1.6 | 6.51 | 6.29 | 0.22 |

**6.1.2 Simulation Results: (QAM and N=128)**

The simulation results are now shown for QAM modulation and no. of subscribers, N=128. In the previous method, PAPR distribution for different CR value is shown in figure 6 (a). At CCDF =$10^{-1}$, the unclipped signal value is 13.11 dB and others values for different CR are tabulated in the table 3. In the proposed method, simulation shows the reduction of PAPR is shown for different CR values in figure 6(b). At CCDF =$10^{-1}$, the unclipped signal value is 13.65 dB and others values for different CR are tabulated in the table 3.

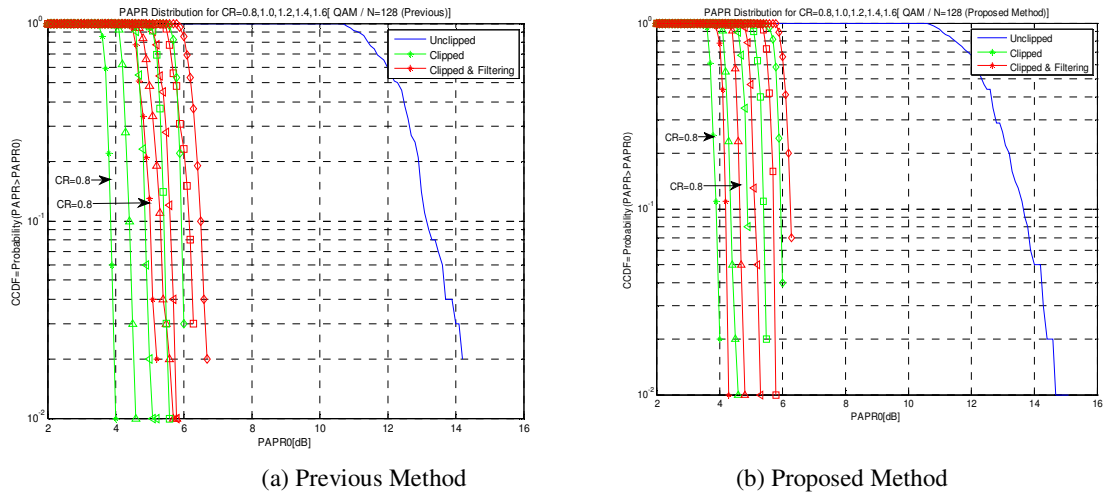

(a) Previous Method                          (b) Proposed Method

Figure 6. PAPR Distribution for CR=0.8,1.0,1.2,1.4,1.6  [QAM and N=128]

The previous method and proposed method PAPR distribution values for different CR values are tabulated and differences are shown in table 3.  From table 3, it is clearly observed that the proposed method reduces PAPR with respect to previous work analysis [7] for QAM and N=128 also. So, proposed method works on efficiently for both QPSK & QAM.

Table 3. Comparison of Previous with Proposed Method for PAPR value [QAM and N=128]

| CR value | PAPR value (dB) (Previous) | PAPR value (dB) (Proposed) | Improvement in PAPR value (dB) |
|---|---|---|---|
| 0.8 | 4.97 | 4.21 | 0.76 |
| 1.0 | 5.25 | 4.65 | 0.60 |
| 1.2 | 5.67 | 5.11 | 0.56 |
| 1.4 | 6.09 | 5.71 | 0.38 |
| 1.6 | 6.51 | 6.27 | 0.24 |

Now, if we compare the values for different CR values in case of QPSK & QAM to show the effect of modulation on proposed filter design, it is observed that for the same number of subscribers (N=128) & low CR=0.8, there is no differences between using QAM & QPSK. But, with the increasing value of CR, QAM provides less PAPR than QPSK. So, for high CR, QAM is more suitable than QPSK in case of proposed filter.

## 6.2. Simulation Results for BER Performance

The clipped & filtered signal is then passed through the AWGN channel and BER are measured for both previous & proposed methods. We have also simulated the analytical BER curve that is shown in the curve. Figure 7 and figure 8 show the BER performance for QPSK and QAM with N=128. It is seen from these figures that the BER increases as the CR decreases.

### 6.2.1 Simulation Results: (QPSK and N=128)

Now, for QPSK & N=128 with all other same data mentioned in table 1, both previous and proposed methods are executed and resulted graphs are shown in figure 7(a) & figure 7(b) respectively. It is observed from these two figures that BER increases slightly in proposed method with respect to previous method for all same values of CR.

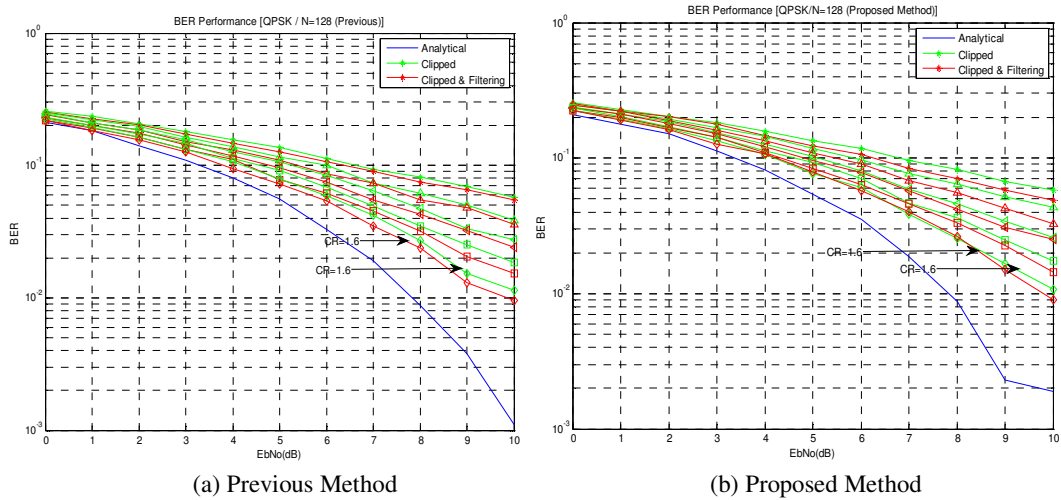

(a) Previous Method     (b) Proposed Method

Figure 7. BER Performance [QPSK and N=128]

The measured BER value at 6 dB point is tabulated in table 4.

Table 4. Comparison of BER value for Previous & Proposed Method [QPSK and N=128]

| CR value | BER value (Previous) | BER Value (Proposed) | Difference in BER value |
|---|---|---|---|
| 0.8 | 0.0752 | 0.10631 | -0.03111 |
| 1.0 | 0.0616 | 0.09012 | -0.02852 |
| 1.2 | 0.0492 | 0.07846 | -0.02926 |
| 1.4 | 0.0411 | 0.06358 | -0.02248 |
| 1.6 | 0.0339 | 0.05748 | -0.02358 |

BER performance is measured and compared in both the table 4 (QPSK) & table 5(QAM) which shows different CR values for both previous & proposed method in case of same parameter value.

From table 4, it is observed that, for CR values (0.8,1.0,1.2,1.4 & 1.6) , the difference magnitude between previous & proposed method are 0.03111,0.02852,0.02926,0.02248 & 0.02358 respectively in QPSK. These BER degradations are acceptable as these are very low values.

### 6.1.2 Simulation Results: (QAM and N=128)

Again, for QAM & N=128 with all other same data mentioned in table 1, both previous and proposed methods are executed and resulted graphs are shown in figure 8(a) & figure 8(b) respectively.

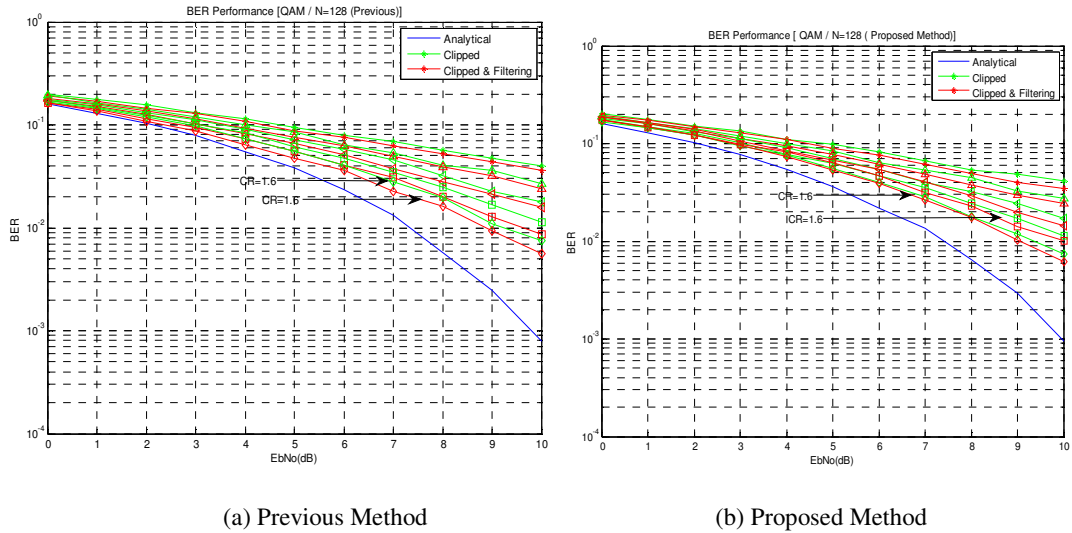

(a) Previous Method  (b) Proposed Method

Figure 8.  BER Performance [QAM and N=128]

The measured BER value at 6 dB point is tabulated in table 5.

Table 5. Comparison of BER value for Previous & Proposed Method [QAM and N=128]

| CR value | BER value (Previous) | BER Value (Proposed) | Difference in BER value |
|---|---|---|---|
| 0.8 | 0.07602 | 0.07535 | 0.00067 |
| 1.0 | 0.06256 | 0.06098 | 0.00158 |
| 1.2 | 0.05091 | 0.05433 | -0.00342 |
| 1.4 | 0.04028 | 0.04631 | -0.00603 |
| 1.6 | 0.03642 | 0.04211 | -0.00569 |

From table 5, it is observed that BER increases slightly in proposed method with respect to previous method for high value of CR= 1.2,1.4,1.6 but in low value of CR=0.8 & 1.0, QAM provides better result ( less BER degradation) in case of proposed method.

## 6. CONCLUSION

In this paper, performances are compared for two different types of amplitude clipping & filtering based PAPR reduction techniques have been analyzed. We ran the simulation for our two techniques, i.e: previous and proposed upon QPSK & QAM. In the new proposed method, it is observed from that for the same number of subscribers (N=128) & low CR=0.8, there is no differences between using QAM & QPSK. But, with the increasing value of CR, QAM provides less PAPR than QPSK. So, for high CR, QAM is more suitable than QPSK in case of proposed filter. In case of BER, with gradual increasing of CR values, the differences of BER values for QPSK become decreasing. But, it is also noticed that BER increases slightly in proposed method with respect to previous method for high value of CR= 1.2,1.4,1.6 but in low value of CR=0.8 & 1.0, QAM provides better result ( less BER degradation) in case of proposed method. In this simulation, ideal channel characteristics have been assumed. In order to estimate the OFDM system performance in real world, multipath Rayleigh fading would be a major consideration in next time. The increase number of subscribers (N) & higher modulation parameters could be another assumption.

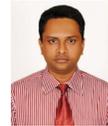

**Authors**

**Md. Munjure Mowla** is now working as an Assistant Professor in Electronics & Telecommunication Engineering department of Rajshahi University Engineering & Technology (RUET) since November 2010. He has completed M.Sc Engineering degree in Electrical & Electronic Engineering (EEE) from RUET in May 2013. He has four years telecom job experience in the companies like operators, vendors, ICX etc of Bangladesh telecom market. Mr. Mowla has published several international journals as well as conference papers and three books. He is a member of IEEE, ComSoc (IEEE), Institutions of Engineers, Bangladesh (IEB) and Bangladesh Electronics Society (BES). His research interest includes advanced wireless communication including LTE, LTE-Advanced, green communication, smart grid communication etc.

**Md. Yeakub Ali** has completed B.Sc Engineering degree in Electronics & Telecommunication Engineering (ETE) from Rajshahi University Engineering & Technology (RUET) in January 2014. Now, he is working as an engineer in a telecommunication company. His research area includes Wireless & Mobile Communication, Satellite Communication & Radar.

**Rifat Ahmmed Aoni** has completed B.Sc Engineering degree in Electronics & Telecommunication Engineering (ETE) from Rajshahi University Engineering & Technology (RUET) in September 2012. Now, he is doing M.Sc (EEE) in University of Malaya, Malaysia. His research area includes Photonics, Optical Fiber etc.